\documentclass[12pt,preprint]{aastex}

\slugcomment{to appear in the Astrophysical Journal}

\begin{document}

\title{A Search for Dense Molecular Gas in High Redshift Infrared-Luminous
Galaxies}

\author{C. L. Carilli$^1$, P. Solomon$^2$, P. Vanden Bout$^3$, F.
  Walter$^1$, Alexandre Beelen$^4$, Pierre Cox$^4$, F. Bertoldi$^5$, K.
  M. Menten$^5$, Kate G. Isaak$^6$, C.J. Chandler$^1$, A. Omont$^7$}

\affil{$^1$National Radio Astronomy Observatory, PO Box O, Socorro, NM 87801,
  USA}
\affil{$^2$Department of Physics and Astronomy, SUNY at Stony Brook,
Stony Brook, NY 11794, USA}
\affil{$^3$National Radio Astronomy Observatory, 520 Edgemont Road,
Charlottesville, VA 22903, USA}
\affil{$^4$Institute d'Astrophysique Spatiale, Universit\'e de Paris XI, 
Orsay F-91405, France}
\affil{$^5$Max-Planck Instit\"ut f\"ur Radioastronomie, Auf dem
  H\"ugel 69, Bonn D-53121, Germany}
\affil{$^6$School of Physics \& Astronomy, University of Wales - 
Cardiff, Cardiff CF24 3YB, UK}
\affil{$^7$Institute d'Astrophysique de Paris, CNRS, \& Universit\'e, 
98 bis bd. Arago, Paris F-75014, France}

\email{ccarilli@nrao.edu}

\begin{abstract}

We present a search for HCN emission from four high redshift far
infrared (IR) luminous galaxies. Current data and models suggest that
these high $z$ IR luminous galaxies represent a major starburst phase
in the formation of spheroidal galaxies, although many of the sources
also host luminous active galactic nuclei (AGN), such that a
contribution to the dust heating by the AGN cannot be precluded.  HCN
emission is a star formation indicator, tracing dense molecular
hydrogen gas within star-forming molecular clouds (n(H$_2$) $\sim
10^5$ cm$^{-3}$).  HCN luminosity is linearly correlated with IR
luminosity for low redshift galaxies, unlike CO emission which can
also trace gas at much lower density.  We report a marginal detection
of HCN (1-0) emission from the $z=2.5832$ QSO J1409+5628, with a
velocity integrated line luminosity of $L_{\rm HCN}'=6.7\pm2.2
\times10^{9}$ K km s$^{-1}$ pc$^2$, while we obtain 3$\sigma$ upper
limits to the HCN luminosity of the $z=3.200$ QSO J0751+2716 of
$L_{\rm HCN}'=1.0\times10^{9}$ K km s$^{-1}$ pc$^2$, $L_{\rm
HCN}'=1.6\times10^{9}$ K km s$^{-1}$ pc$^2$ for the $z= 2.565$
starburst galaxy J1401+0252, and $L_{\rm HCN}'=1.0\times10^{10}$ K km
s$^{-1}$ pc$^2$ for the $z = 6.42$ QSO J1148+5251.  We compare the HCN
data on these sources, plus three other high-$z$ IR luminous galaxies,
to observations of lower redshift star-forming galaxies. The values of
the HCN/far-IR luminosity ratios (or limits) for all the high $z$
sources are within the scatter of the relationship between HCN and
far-IR emission for low $z$ star-forming galaxies.  These observations
are consistent with dust heating by a massive starburst in these
systems, with two important caveats.  First, about half the
measurements are strictly upper limits to the HCN luminosities. And
second, the IR spectral energy distributions for most of the high $z$
sources are well constrained only on the Rayleigh-Jeans side of the
thermal dust peak.  We also present a spatially resolved 42 GHz
continuum image of the gravitational lens J0751+2716.

\end{abstract}


\section{Introduction}

Blind surveys, and targeted observations of known sources
(quasi-stellar objects and radio galaxies), with mJy sensitivity at
submillimeter (submm) wavelengths have revealed a population of IR
luminous galaxies at high redshift, with luminosities $\ge 10^{12}$
L$_\odot$, placing them in the category of Ultraluminous Infrared
Galaxies (ULIRGs; see reviews by Blain et al. 2003; Sanders \& Mirabel
1996). Current models suggest that this population may represent the
formation of large spheroidal galaxies at $z > 2$ (Blain et al.
2003).  If star formation dominates the dust heating in the high-$z$
systems, then the implied star formation rates (up to $10^3$ M$_\odot$
year$^{-1}$) are such that a significant fraction of the stars in a
spheroidal galaxy could be formed in 10$^8$ years. However, in many
cases a contribution to the IR luminosity from dust heated by an AGN
cannot be precluded (Andreani et al. 2003).

An important observation in this regard has been the detection of
giant reservoirs of molecular gas via CO emission lines, with gas
masses $\ge 10^{10}$ M$_\odot$, providing the requisite material for
star formation (Carilli et al. 2004a).  However, CO can be excited at
relatively low densities -- with a critical density for excitation of
only $\rm n(H_2) \sim 10^3$ cm$^{-3}$ for the lower $J$ transitions,
where $J$ is the angular momentum quantum number. Hence it is a good
tracer of the total molecular gas content of galaxies, but it is a
relatively poor tracer of the denser gas directly involved in massive
star formation. This fact is accentuated by the non-linear relation
between IR luminosity and CO luminosity in star-forming galaxies, with
IR luminosity increasing as CO luminosity roughly to the power 1.7
(Gao \& Solomon 2004a,b).  The non-linear increase in IR luminosity
with increasing CO luminosity has been interpreted as an increase in
the star formation efficiency, defined as the ratio of star formation
rate to total gas mass, with increasing star formation rate (Solomon
et al. 1992; Gao \& Solomon 2004b).

Local (z $< 0.3$) ULIRGs show strong HCN emission (Solomon, Downes, \&
Radford 1992; Gao \& Solomon 2004a).  HCN emission traces much denser
gas (critical density for excitation of the lower order transitions
$\rm n(H_2) \sim 10^5~\rm cm^{-3}$; Evans 1999) than CO emission due
to a higher dipole moment.  In the Milky Way, very strong HCN emission
is found in molecular cloud cores, the sites of star formation, and
not in the more massive but less dense cloud envelopes (Helfer \&
Blitz 1997). Strong HCN emission is therefore an indicator of active
star formation. This fact has been demonstrated for external galaxies
by the tight, linear correlation between IR and HCN luminosity for
star-forming galaxies over a wide range of IR luminosity (10$^9$ to
10$^{12}$ L$_\odot$; Gao \& Solomon 2004a,b).  For the most luminous
infrared galaxies the HCN line luminosities range from 1/4 to 1/10
that of the CO luminosity, as compared with ordinary spiral galaxies
where the ratio is typically 1/25 to 1/40. The fact that the ratio of
IR luminosity to HCN luminosity is the same in ULIRGs as in lower
luminosity galaxies suggests that ULIRGs, like the lower luminosity
galaxies, are primarily powered by star formation, and that the HCN
luminosity is a good measure of the mass of actively star-forming
cloud cores (Gao \& Solomon 2004b).  In essence, the star formation
which is responsible for the IR emission has a rate that is
linearly proportional to the mass of dense (ie. HCN-emitting)
molecular gas, but not to the total molecular gas, as traced by CO.

The potential of HCN observations as a star formation diagnostic at
high redshift was demonstrated recently with the detection of HCN (1-0)
emission from the Cloverleaf quasar at $z=2.6$ (H1413+117) using the
Very Large Array (VLA) (Solomon et al. 2003).  These observations have
proven instrumental in the physical interpretation of the
starburst-AGN connection in this system (section 5). More
recently, HCN (1-0) emission has been detected from IRAS~F10214+4724
using the Green Bank Telescope (Vanden Bout, Solomon, \& Maddalena
2004), which may also be an AGN-starburst system. The starburst-AGN
connection has taken on new importance with the discovery of the black
hole mass -- bulge mass relation, suggesting a ``causal connection
between the formation and evolution of the black hole and the bulge''
(Gebhardt et al. 2000).

In this paper we present a search for HCN emission from four high
redshift galaxies using the VLA.  The sensitivity of these
observations is such that we could detect the sources at the level
seen for low $z$ galaxies, given their IR luminosities.  We combine
these results with three sources from the literature, and discuss the
relationship between HCN and IR luminosity for high $z$ ULIRGs, and
possible consequences for star formation.  We assume a standard
concordance cosmology throughout, with H$_o = 70$ km s$^{-1}$
Mpc$^{-1}$, $\Omega_M = 0.3$, and $\Omega_\Lambda = 0.7$.

\section{Sources}

The four sources were selected for high IR luminosity ($L_{\rm IR} >
10^{12}$ L$_\odot$), strong CO emission, and therefore a precise
redshift determination which allows for study of HCN lines with the
VLA, with the obvious requirement that the HCN lines redshift into one
of the VLA receiver bands.  Three were originally discovered as
optical or radio quasars, while the fourth is a submm-selected galaxy,
with no evidence for an AGN.  In the analsys below we also include
three recent high redshift HCN detections from the literature:
IRAS 10214+4724 at $z=2.286$ (Solomon \& Vanden Bout 2004),
BR 1202+0725 at $z=4.694$ (Isaak et al. 2004), and H1413+117 (the 
'cloverleaf') at $z=2.558$ (Solomon et al. 2003). All three
of the sources are characterized by strong thermal IR emission,
and show evidence for an AGN in their optical spectra. 

A key point in our analysis is the derivation of the far-IR
luminosities. Most of the values quoted herein (Table 2) were recently
computed by Beelen et al. (2004, in prep) using single temperature
grey body models fit to all the current photometric measurements.  For
most of the sources the observational data are limited to the
Rayleigh-Jeans side of the spectral energy distribution (SED), with
only marginal sampling of the thermal dust peak (Benford et al. 1999,
Priddey \& McMahon 2001, Beelen et al. 2004, in prep).  Hence, the
dust temperatures are poorly constrained.  The mean source SED of
high-$z$ sources in the Beelen et al. analysis has $T_{\rm dust} \sim
50 \, K$ and a dust emissivity index of $\beta \sim 1.6$,
characteristic of ULIRGs at low redshift (Sanders \& Mirabel 1996).
However, for most of the sources the poor sampling of the dust
emission peak implies that we cannot rule out a lower mass, but higher
temperature dust component ($\ge 100$ K) which would dominate in the
(rest-frame) mid-IR ($\sim 10$ to 40$\mu$m), perhaps heated by an
AGN. This hot mid-IR component could dominate the total IR emission
(ie. integrated from 1 to 1000$\mu$m), but contribute only a fraction
of ($< 30\%$) to the far-IR luminosity ($\sim 40$ to 120 $\mu$m).  For
this analysis we have derived the luminosity by integrating over a
modified black body fitted to the rest-frame far-IR SEDs of each
object, and we set a value $\beta = 1.5$ when not enough photometric
data points are available to fit both $\beta$ and T.  The one notable
exception from the literature is the Cloverleaf quasar (H1413+117),
for which the IR SED is well sampled in frequency (Weiss et
al. 2003). We discuss this issue in more detail in section 5.

{\bf MG~0751+2716:} This is a strongly lensed (magnification factor
17), radio-loud QSO at $z=3.200$ with complex structure in the image
plane on a scale of 1$''$ (Lehar et al 1997; Barvainis et
al. 2002). It is a IR luminous galaxy, with an (apparent) IR
luminosity of $2\times10^{13}$ L$_\odot$, and shows strong CO
emission, with a velocity integrated CO (4-3) flux of 6.0 Jy km
s$^{-1}$ (Barvainis et al. 2002).  The radio source shows a falling
spectrum, with a flux density at 15 GHz of 48 mJy and a spectral index
between 8 and 15 GHz of $\alpha = -1.2$ (Lehar et al. 1997).

{\bf J1148+5251:} The source SDSS J1148+5251 is the most distant QSO
known, at $z = 6.42$ (Fan et al. 2003). Thermal emission from warm
dust was detected from J1148+5251 at (sub)mm wavelengths (Bertoldi et
al. 2003b; Robson et al. 2004; Beelen et al. 2004, in prep), with an
implied rest frame IR luminosity of $2.7\times 10^{13}$ L$_\odot$.
Multiple transitions of CO emission have been detected from this
galaxy, with a velocity integrated CO (6-5) flux of 0.73 Jy km
s$^{-1}$ (Bertoldi et al. 2003b; Walter et al. 2003).  Non-thermal
radio continuum emission has also been detected at 1.4 GHz from
1148+5251, with a flux density of $55\pm12\ \mu$Jy, consistent with
the radio-IR correlation for star-forming galaxies (Carilli et
al. 2004b). There is no evidence for strong gravitational lensing
(i.e., multiple imaging) of this source in high resolution optical and
radio images (Carilli et al. 2004b).

{\bf SMM~1401+0252:} SMM~1401+0252 at $z = 2.565$ was detected in the
submm survey of cluster fields of Ivison et al. (2001), with an
apparent IR luminosity of $7.5\times10^{12}$ L$_\odot$, and multiple
CO transitions have been detected from this galaxy (Frayer et
al. 1999; Downes \& Solomon 2003).  The source is gravitationally
lensed by a foreground cluster, and perhaps by a galaxy along the line
of sight. Swinbank et al. (2004) estimate a magnification factor for
the CO of 5, and Downes \& Solomon (2003) derive a limit to the CO
source size $< 2"$.  Optical spectra are consistent with a
star-forming galaxy, showing no evidence for an AGN (Ivison et
al. 2000). The source is detected at 1.4 GHz with a flux density
consistent with the radio-IR correlation for star-forming galaxies
(Ivison et al. 2001).

{\bf J1409+5628:} This optically selected QSO at $z=2.5832$ is the
most luminous IR source in the sample of $z=2$ to 3 QSOs of Omont et
al.  (2003), with $L_{IR} = 3.3\times 10^{13} \, L_\odot$.  It also
shows strong CO emission, with a velocity integrated CO (3-2) flux of
2.3 Jy km s$^{-1}$ (Beelen et al. 2004; Hainline et al. 2004).  VLA
observations show an unresolved 1 mJy source at 1.4 GHz with a
spectral index between 1.4 and 5 GHz of $-0.75$, consistent with the
radio-IR correlation for star-forming galaxies (Petric et al. 2004).
High resolution imaging with the VLBA shows a resolved radio source
with an intrinsic brightness temperature (at 8 GHz) of 10$^5$ K, again
consistent with a star-forming galaxy (Beelen et al. 2004). There is
no evidence for strong gravitational lensing of J1409+5628.

\section{Observations}

Three of the sources were observed in the C configuration of the VLA
(maximum baseline length 3 km), while J1148+5251 was observed in the D
configuration (maximum baseline 1 km).  The observational parameters
are given in Table 1. Each observing day entailed between 6 and 8
hours on-source time.  Amplitude calibration was performed using 3C286,
while fast switching phase calibration was employed on timescales of 3
minutes, as well as dynamic scheduling to ensure good weather. On all
days the phase stability was excellent.

For two of the sources the HCN (1-0) transition was observed in the 22
GHz band of the VLA using spectral line mode with two polarizations, 7
spectral channels per polarization, and 6.25 MHz per channel (= 78 km
s$^{-1}$).  Based on the CO redshifts, the expected line centers for
J1409+5628 and SMM~1401+0252 are 24.7354 and 24.8617 GHz,
respectively. Due to VLA tuning restrictions with a 50 MHz
bandbass, we centered the observations at the closest allowed
frequencies of 24.7351 and 24.8649 GHz. For J1409+5628 each day also
included one hour in standard continuum mode at 22 GHz to obtain a
sensitive limit on the continuum emission.

For MG~0751+2716 we observed the HCN (2-1) transition in the 43 GHz
band using continuum mode with two IFs of two polarizations and 50 MHz
bandwidth each ($\sim 350$ km s$^{-1}$).  The continuum mode was
selected due to severe spectral restrictions in the VLA correlator
when observing with a 50 MHz bandwidth. The mode chosen optimizes
sensitivity to a line of width $\sim 300$ km s$^{-1}$, but sacrifices
spectral resolution. Based on the CO redshift, the expected line
center is 42.208 GHz, but again, due to VLA tuning restrictions we
centered the line IF at the closest allowed frequency of 42.215
GHz. The off-line IF was sequentially tuned $\pm 150$ MHz above and
below the line frequency to obtain a high quality image of the
continumm source.

For the highest redshift source, J1148+5251, we used two IFs to
observe simultaneously the HCN (2-1) transition at 23.8929~GHz and the
HCO$^+$ (2-1) transition at 24.0430~GHz. Each IF had two polarizations
and seven spectral channels with a channel width of 3.125~MHz
(= 39 km s$^{-1}$). HCO$^+$ is also a dense gas
indicator, with a critical density for excitation similar to HCN 
(Evans 1999). 

\section{Results}

{\bf J1409+5628:} This source is marginally detected in the radio
continuum at 22.5 GHz with S$_{22} = 67\pm 23\ \mu$Jy.  The implied
spectral index between 1.4 and 22.5 GHz is $-1.0 \pm 0.13$, consistent
with the spectral index measured between 1.4 and 5 GHz (Petric et
al. 2004).  The CLEAN components from the continuum image were
subtracted from the line data, and the resulting HCN (1-0) spectrum is
shown in Figure 1. Zero velocity corresponds to the CO heliocentric
redshift.  The spectrum shows a possible detection of HCN emission in
at least two channels. The contour image of the average of these two
channels is shown in Figure 2.  Gaussian fitting to the line profile
results in a peak of $82\pm30\ \mu$Jy, a FWHM = $177\pm 80$ km
s$^{-1}$, and a central velocity of $-43 \pm 30$ km s$^{-1}$ relative
to the CO redshift. Overall, we feel this is a marginal detection of
HCN emission from J1409+5628, and given possible uncertainties in the
continuum subtraction, and the relatively low signal-to-noise ratio
per channel, we do not consider the low FWHM (for comparison, the FWHM
of the CO (3-2) line = 311 km s$^{-1}$; Beelen et al. 2004), or the
velocity offset, relative to the CO line to be significant.  We
calculate the HCN velocity integrated line luminosity in K km s$^{-1}$
pc$^{2}$ using equation 1 from Solomon, Radford, \& Downes (1992), and
using the nominal values from the Gaussian fitting to the line
profile. The HCN line luminosity is listed in column 5 of Table 2.

{\bf MG~0751+2716:} The radio continuum image of MG~0751+2716 is shown
in Figure 3.  The continuum image shows the multiple structure
expected for this complex lensed source (Lehar et al. 1997), with a
total flux density of 13.2 mJy, consistent with a spectral index of
--1.2 extending from 8 GHz to 42 GHz.  The line image (after continuum
subtraction) shows no emission with an rms level of 0.1 mJy.  We have
also convolved the image to 2$''$ resolution to search for extended
emission (recall that the source has structure due to gravitational
lensing on a scale of $\sim 1''$), and no emission is seen to an rms
level of 0.3 mJy. We calculate an upper limit to the HCN line
luminosity (Table 2) assuming an upper limit of 0.3 mJy for a channel
width of 350 km s$^{-1}$.

{\bf SMM~1401+0252:} No emission is seen from this source with an rms
per channel of 44 $\mu$Jy. Averaging over all channels sets a
3$\sigma$ limit to the continuum emission of 50 $\mu$Jy. The CO line
FWHM is 200 km s$^{-1}$. Averaging over the three central channels of
the HCN spectrum (= 234 km s$^{-1}$) leads to a 3$\sigma$ limit of 76
$\mu$Jy, with an implied HCN line luminosity as given in Table 2.

{\bf J1148+5251:} No HCN or HCO+ line emission is detected from
J1148+5251, with an rms per 39 km s$^{-1}$ channel of 70
$\mu$Jy. Summing over the band (273 km s$^{-1}$) gives a 3$\sigma$
limit to the integrated HCN or HCO+ line emission of 90 $\mu$Jy, or a
limit to the line luminosity as given in Table 2.  Summing both the
HCN and HCO+ data gives a 3$\sigma$ upper limit to the continuum
emission at 24 GHz of 70 $\mu$Jy.

Two continuum sources are detected in the field of J1148+5251, located
about 1$'$ northeast and southwest of the QSO, with flux densities of
4.9 and 4.1 mJy at 24 GHz. These sources have been detected previously
at 1.4 GHz, and have been discussed at length in Carilli et
al. (2004b), with the SW source corresponding to the core of a radio
galaxy at $z = 0.05$, and the NE source having a point source optical
counterpart of unknown redshift.

\section{Discussion}

We summarize the HCN results for these four high redshift IR luminous
galaxies in Table 2, along with results for three other high redshift
sources from the literature (Solomon et al. 2003; Isaak et al. 2004;
Solomon \& Vanden Bout 2004).  We assume constant
brightness temperature when extrapolating from higher order
transitions to the 1--0 transition, ie. $L'$ is independent of
transition.  Multiple transition CO emission line studies of a few
high $z$ sources, including BR~1202+0725 in Table 2, show that this is
a reasonable assumption, at least up CO(4--3) (Carilli et al. 2002).
Whether this is also true for the HCN(2--1) to HCN(1--0) transitions
remains to be verified.  Column 7 gives the gravitational lens
magnification factor used to correct to intrinsic luminosity.

Figure 4 shows the relationship between far-IR luminosity and HCN line
luminosity for the seven sources in Table 2, plus the local galaxy
samples ($z < 0.1$) of Gao \& Solomon (2004a) and Solomon, Downes, \&
Radford (1992).  For the Gao \& Salomon (2004a) sample, we re-derived
the far-IR luminosities of the sources using the flux densities listed
in the revised IRAS bright galaxy catalog and again integrating over a
modified black body fitted to the rest-frame far-infrared SEDs of each
object. The solid line shows the relationship between the far-IR and
HCN luminosity of the Gao \& Solomon (2004a) and Solomon, Downes, \&
Radford (1992) samples: $\log L_\mathrm{FIR} = 1.09 \log
L_\mathrm{HCN} + 2.0$. The fact that the power-law index is close to
unity ($1.09\pm0.02$) implies a nearly linear relationship between FIR
and HCN luminosity for the low $z$ galaxy sample.  Overall, the high
$z$ sources fall within the scatter of the low $z$ source
relationship, although four of the cases are strictly HCN upper
limits. Thus far, the data are consistent with the hypothesis that the
same dust heating mechanism operates in both low and high $z$ sources,
and in the case of the low $z$ sources this mechanism has been shown
to be star formation (Gao \& Solomon 2004b).

As an example, consider J1409+5628.  A rough estimate of the dense gas
mass can be derived from the relation $M_{\rm dense} = \alpha_{\rm
HCN} L_{\rm HCN}')$, where $\alpha_{\rm HCN} \sim 7$ M$_\odot$ (K km
s$^{-1}$ pc$^{2}$)$^{-1}$ assuming the gas is warm ($\sim 50$K; Gao \&
Solomon 2004b), leading to a dense gas mass of $5\times10^{10}$
M$_\odot$. Beelen et al. (2004) estimate a (total) star formation
rate for the J1409+5628 galaxy of $\sim 5000$ M$_\odot$ year$^{-1}$
from the far-IR and radio continuum luminosities.  Hence, the dense
gas depletion timescale is of order $10^7$ years.  The implication is
that the starburst in J1409+5628 is relatively short-lived, unless the
dense gas can be re-supplied on timescales $\le 10^7$ years.

However, we should emphasize that all the high $z$ sources (including
the upper limits) in Figure 4 fall on the low side of the relation set
by the low $z$ samples in terms of their HCN luminosity.  This trend
may imply some contribution to dust heating by another source, such as
an AGN, or a possible breakdown in the low $z$ relationship between
HCN and IR luminosities at the highest luminosities (or at high
redshift).  Clearly a larger sample of high $z$ IR luminous galaxies
with sensitive HCN observations are required to test these
possibilities.

For the high $z$ sources, the one case in Table 2 for which the rest
frame IR SED is well sampled across the dust peak is the Cloverleaf
quasar (H1413+117 in Table 2; Weiss et al. 2003).  Weiss et al. (2003)
find the IR emission can be decomposed into a warm component (115 K)
that dominates the mid-IR emission, and a cooler component (50 K) that
dominates the far-IR emission. The cool component also dominates the
dust mass.  In their study of the HCN emission from the Cloverleaf,
Solomon et al. (2003) show that the HCN/IR ratio for the cool
component follows the low $z$ star-forming galaxy relation, and from
this they propose that 22$\%$ of the total IR luminosity
(corresponding to the cooler component) results from dust heated by
star formation, while the hotter component is dust heated by the AGN.
In Figure 4 we have only included the cool component in the IR
luminosity for the Cloverleaf.  It is possible that some of the other
sources in Table 2, in particular those harboring known AGN, have
similar hot dust components. Future sensitive observations at higher
frequencies ($> 350$ GHz and above) are required to test this
hypothesis.

The observations presented herein highlight the difficulty in
detecting thermal molecular transitions other than CO from high
redshift IR luminous galaxies with current instruments. The improved
spectral capabilities of the Expanded Very Large Array, and the wider
total redshift range covered by the
receivers\footnote{http://www.aoc.nrao.edu/evla/}, will help in terms
of detecting low order transitions redshifted to centimeter
wavelengths, although the nominal line sensitivity of the array at
these frequencies will improve by at most a factor two relative to the
current system at 20 to 50 GHz.  The Atacama Large Millimeter Array
(ALMA)\footnote{http://www.alma.nrao.edu/}, will improve the
sensitivity in the millimeter regime by more than an order of
magnitude relative to current instruments. However, ALMA will be
restricted to studying the higher order transitions (3-2 and higher
for $z > 2$) of the most common tracers such as HCN and HCO+.  These
transitions may be sub-thermally excited, as is seen for HCN(J=4-3) in
the Cloverleaf (Guelin, et al. 2004, in prep; Solomon et al. 2003),
due to the very high critical densities for excitation of the higher
levels ($\rm n(H_2) \sim 10^7 ~ cm^{-3}$; Evans 1999).

\acknowledgments

The National Radio Astronomy Observatory is operated by Associated
Universities Inc., under cooperative agreement with the National
Science Foundation. The authors thank the referee for a careful
review of this paper.

\clearpage


\clearpage\newpage

\begin{deluxetable}{ccccccccc}
\tabletypesize{\scriptsize}
\tablecaption{Observational Parameters \label{tbl-1}}
\tablewidth{0pt}
\tablehead{
\colhead{Source} & \colhead{Date} & \colhead{Transition} & 
\colhead{Rest Freq.} &\colhead{Obs. Freq.} & 
\colhead{FWHM} & \colhead{rms} & \colhead{Chan. Width}& 
\colhead{Bandwidth}} 
\startdata
~ & ~ & ~ & GHz & GHz & arcsec & $\mu$Jy & km s$^{-1}$ & MHz \\
\hline 
MG 0751+2716 & April 16, 18, 2004     & HCN(2--1)     & 177.261 & 42.215 & 0.6 
& 100 & 350 & 3x50 \\
J1148+5251 & June  21, 26, 2004     & HCN(2--1)     & 177.261 & 23.892 & 3.5 & 60 
& 39 & 25 \\
J1148+5251 & June  21, 26, 2004     & HCO$^+$(2--1) & 178.375 & 24.043 & 3.5 & 60 
& 39 & 25 \\
SMM 1401+0252 & March 21, 26, 2004     & HCN(1--0)     & 88.632 & 24.865 & 1.0 
& 44 & 78 & 50 \\
J1409+5628 & April 16, 18, 28, 2004 & HCN(1--0)     & 88.632 & 24.735 & 1.1 & 40 
& 78 & 50 \\
\enddata
\end{deluxetable}

\begin{deluxetable}{cccccccc}
\tabletypesize{\scriptsize}
\tablecaption{HCN observations of high $z$ infrared-luminous galaxies 
\label{tbl-2}}
\tablewidth{0pt}
\tablehead{
\colhead{Source} & \colhead{Type} & \colhead{$z^a$} & 
\colhead{$L_{\rm FIR}^{b,c}$} 
& \colhead{$L_{\rm HCN}'^d$} & \colhead{$L_{\rm CO}'^e$} 
& \colhead{Mag$^f$}  & \colhead{References: CO; HCN}}
\startdata
~ & ~ & ~ & $10^{12}$L$_\odot$ & \multicolumn{2}{c}{$10^9$ 
K km s$^{-1}$ pc$^2$} & ~ & ~ \\
\hline 
MG 0751+2716 & QSO & 3.200 & $1.2$ & $< 1.0$ 
& $9.7$ & 17 & Barvainis et al. 2002; This paper \\
IRAS F10214+4724 & AGN & 2.286 & $4.0$ & $1.3\pm0.3$ 
& $9.1$ & 13 & Solomon et al. 1992; Solomon \& 
Vanden Bout 2004 \\
J1148+5251 & QSO & 6.419 & $27$ & $< 10$ 
& $27$ & 1 & Walter et al. 2003; This paper \\
BR 1202+0725 & QSO & 4.693 & $60$ & $< 49$ 
& $89$ & 1 & Carilli et al. 2002; Isaak et al. 2004 \\
SMM 1401+0252 & Gal & 2.565 & $1.5$ & $< 1.6$ 
& $19$ & 5 & Ivison et al. 2001; This paper \\
J1409+5628 & QSO & 2.583 & $33$ & 
$6.7\pm2.2$  & $82$ & 1 & Beelen et al. 2004; 
This paper \\
H1413+117 & QSO & 2.558 & $6.2$ & $3.2\pm0.5$
& $46$ & 11 & Weiss et al. 2004; Solomon et al. 2003 \\
\enddata
\tablecomments{
~$^a$The redshift derived from the CO emission.  \\
~$^b$The intrinsic far-IR luminosities derived as discussed in section 2. \\
~$^c$All the line and continuum luminosities in this Table have been 
corrected for gravitational magnification (column 7). \\
~$^d$ The velocity integrated HCN (1-0) line luminosity (or 3$\sigma$ upper
limits). \\
~$^e$The velocity integrated CO (1-0) line luminosity (or 3$\sigma$ upper
limits). \\
~$^f$The gravitational lens magnification factor. 
}
\end{deluxetable}

\clearpage\newpage

\begin{figure}
\plotone{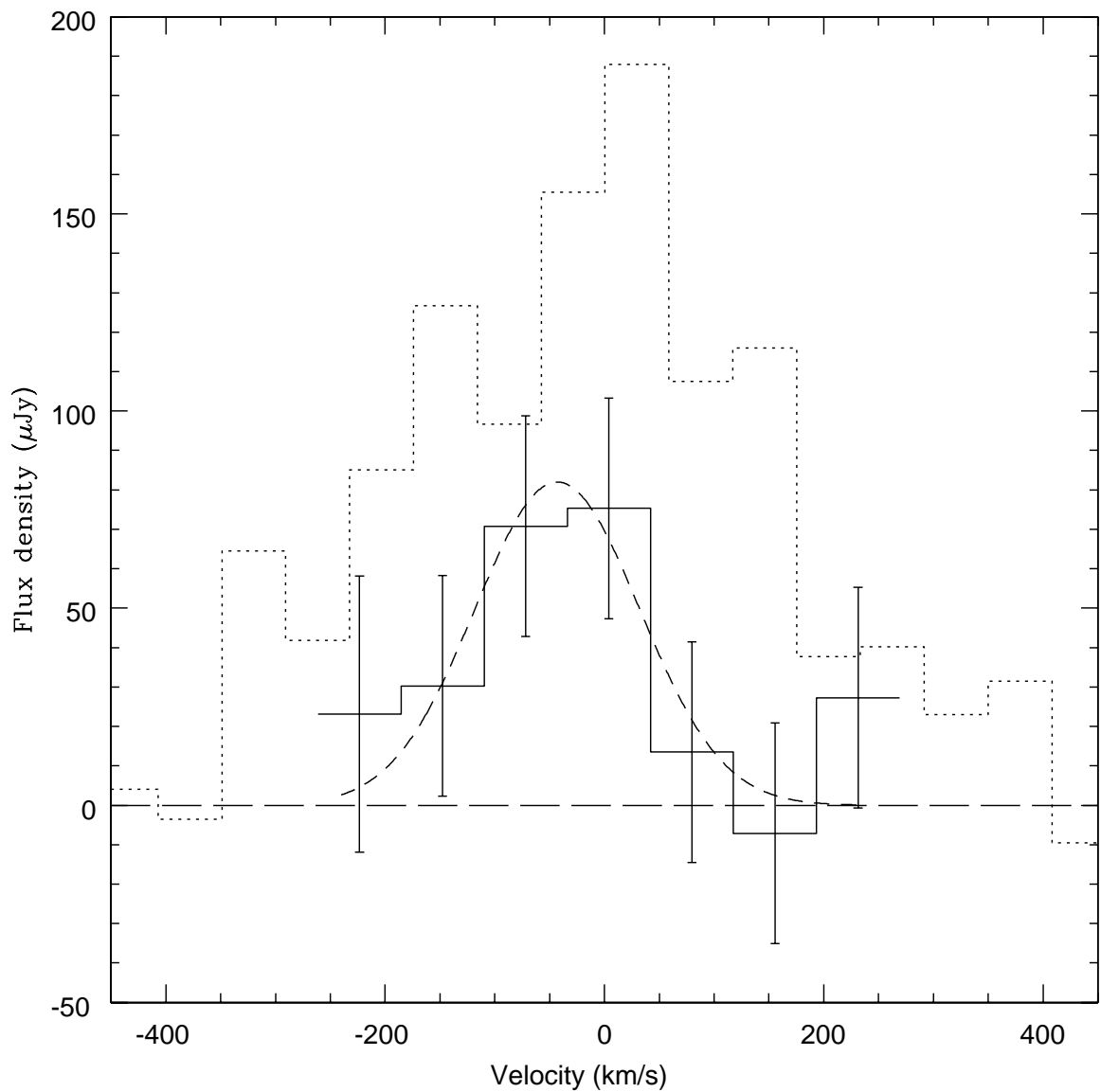}
\caption{VLA spectrum of HCN (1-0) emission from J1409+5628 (solid line plus
points with error bars),
with continuum subtracted as described in section 4.
Zero velocity corresponds to the CO (3-2) redshift of $z =
2.5832$. This spectrum has been hanning smoothed, such that each
channel is not independent. The rms per hanning smoothed channel is 28
$\mu$Jy. The dashed line shows a Gaussian fit to the data with
parameters given in section 4. The dotted line shows the CO (3-2) spectrum
from Beelen et al. (2004), scaled by a factor 1/40. }
\end{figure}

\clearpage\newpage

\begin{figure}
\plotone{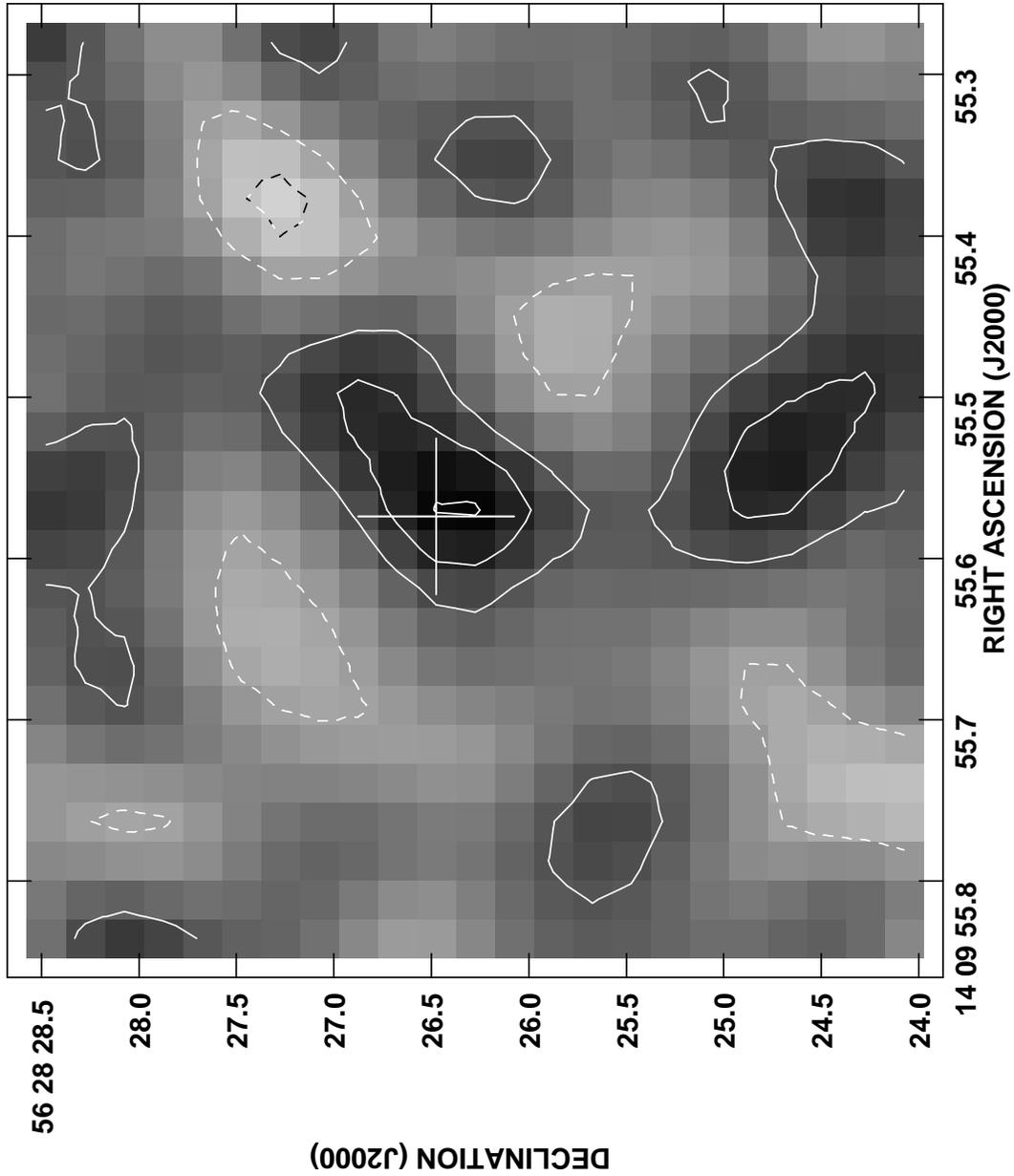}
\caption{VLA image (contours and greyscale) of the average of
the two peak channels containing possible HCN emission 
from  J1409+5628 (see Figure 1). The contour levels are -64,-32, 
32, 64, 96 $\mu$Jy beam$^{-1}$ and the resolution FWHM = $1.1''$.
Negative contours are dashed. The radio QSO position is indicated 
by a cross. The rms noise on this image is 28 $\mu$Jy beam$^{-1}$.
}
\end{figure}

\clearpage\newpage

\begin{figure}
\plotone{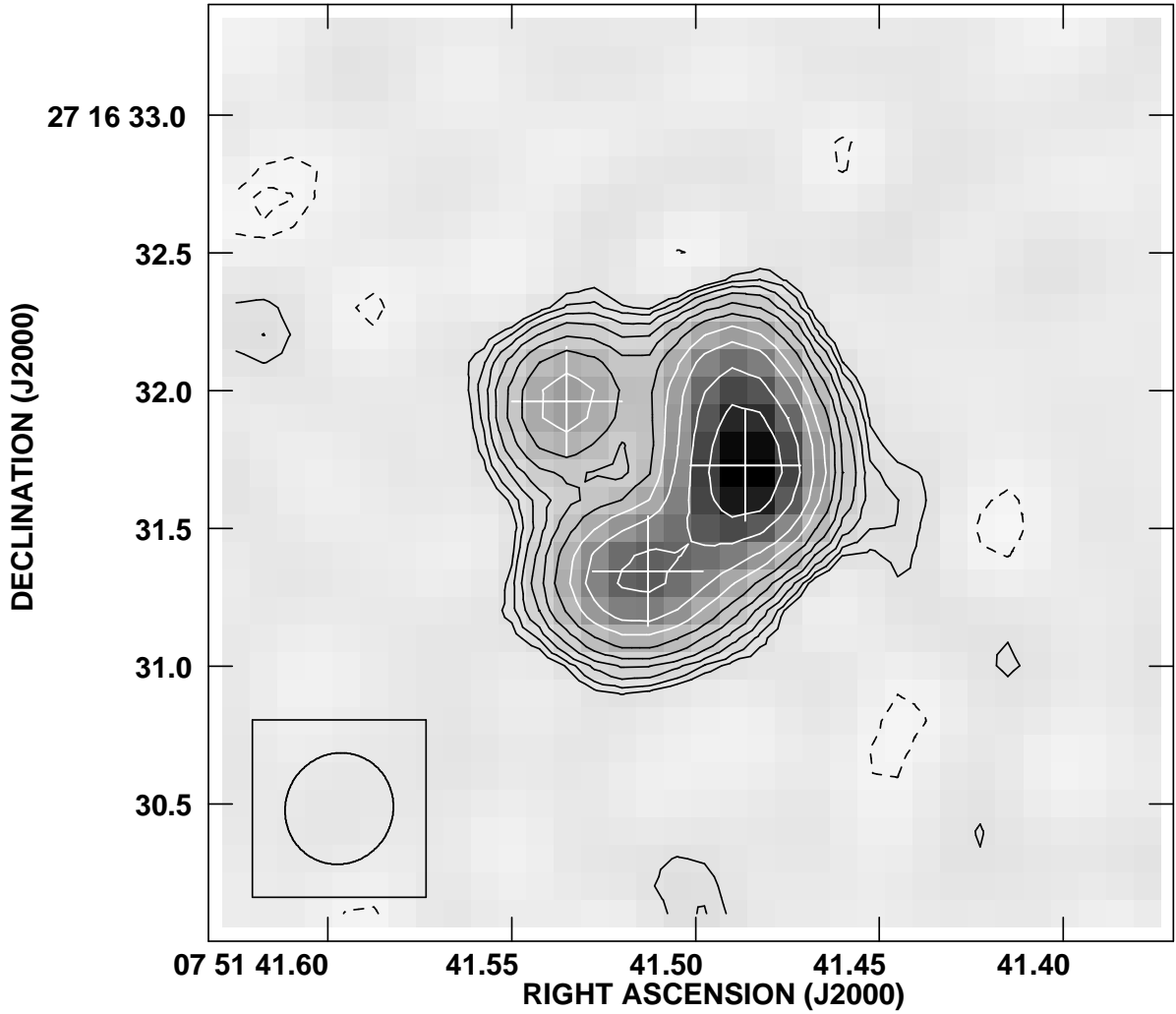}
\caption{VLA image (contours and greyscale) of the radio
  continuum emission from the $z = 3.200$ gravitationally lensed QSO
  MG~0751+2716 at 42.2 GHz.  The contour levels are a geometric
  progress in $\sqrt{2}$, such that two contours corresponds to a
  change in surface brightness by a factor two, starting at 0.17 mJy
  beam$^{-1}$. The FWHM = $0.60''$. The crosses indicate the local
  peak surface brightness positions for the three main lensed radio
  components.  }
\end{figure}

\clearpage\newpage

\begin{figure} 
\epsscale{0.8}
\plotone{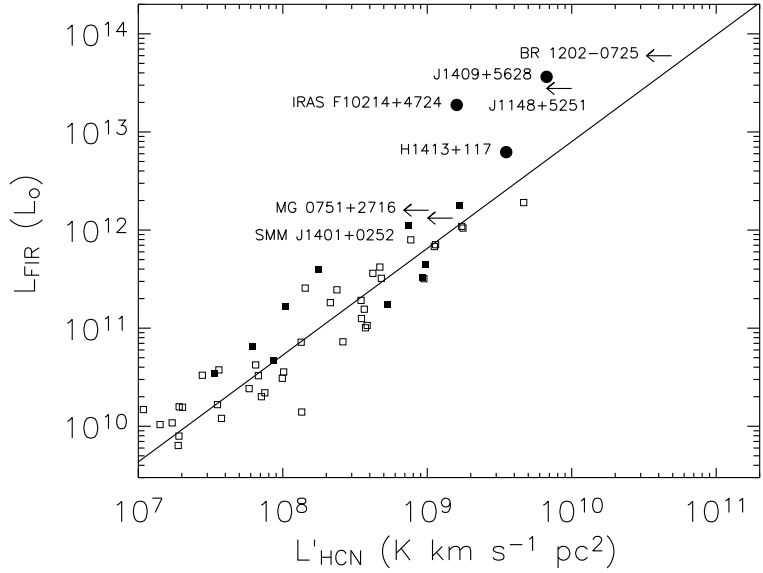} 
\caption{ A comparison of
  velocity integrated HCN line luminosity with far-IR luminosity for a
  sample of low $z$ galaxies (open squares from Gao \& Solomon
  2004a; filled squares from Solomon, Downes, \& 
  Radford (1992)), and for the
  high $z$ ULIRGs in Table 2 (filled circles for detections, or arrows
  indicating 3$\sigma$ upper limits). All are corrected for
  gravitational magnification as per Table 2. 
For the Gao \& Salomon (2004a) sample, we re-derived
the far-IR luminosities of the sources using the flux densities listed
in the revised IRAS bright galaxy catalog and integrating over a
modified black body fitted to the rest-frame far-IR SEDs of each
object. The solid line is the relationship defined by the low $z$ galaxies,
corresponding to: $\log L_\mathrm{FIR} = 1.09 \log
L_\mathrm{HCN} + 2.0$. }
\end{figure}

\clearpage\newpage

\end{document}